\documentclass[final, 5p, times, twocolumn]{elsarticle}
\usepackage{booktabs}
\usepackage{diagbox}

\usepackage{hyperref}
\hypersetup{
  colorlinks=true, % Enable colored text links (disables boxes)
  linkcolor=blue, % Color for internal links
  urlcolor=blue, % Color for external URLs
  filecolor=blue, % Color for file links
  citecolor=blue % Color for citations
}

\usepackage[inkscapelatex=false]{svg}

\usepackage{amssymb}
\usepackage{amsmath}
\usepackage{lipsum}
\makeatletter
\let\old@refstepcounter=
\refstepcounter

\renewcommand{\refstepcounter}[1]{%
\old@refstepcounter{#1}%
\protected@edef\@currentlabel{\textsection\@currentlabel}%
}
\makeatother
\usepackage{xcolor}
\usepackage{soul}
\usepackage[normalem]{ulem}

\journal{Future Gener. Comput. Syst.}

\begin{document}

\begin{frontmatter}



    %\title{AI4EOSC: A federated platform for AI in Science} %% Article title
    \title{AI4EOSC: a Federated Cloud Platform for Artificial Intelligence in Scientific Research}


    \author[ifca,first]{Ignacio Heredia}
    \author[ifca,first]{Álvaro López García\corref{cor1}}
    \ead{aloga@ifca.unican.es, aloga@ifca.es}
    \cortext[cor1]{Corresponding author}

    \author[ifca]{Fernando Aguilar Gómez}
    \author[upv]{Diego Aguirre}
    \author[upv]{Caterina Alarcón Marín}
    \author[kit]{Khadijeh Alibabaei}
    \author[kit]{Lisana Berberi}
    \author[upv]{Miguel Caballer}
    \author[upv]{Amanda Calatrava}
    \author[predictia]{Pedro Castro}
    \author[infn]{Alessandro Costantini}
    \author[lip]{Mario David}
    \author[predictia]{Jaime Díez}
    \author[iisas]{Stefan Dlugolinsky}
    \author[infn]{Giacinto Donvito}
    \author[kit]{Leonhard Duda}
    \author[kit]{Borja Esteban Sanchis}
    \author[ifca]{Saúl Fernandez Tobías}
    \author[ifca]{Andrés Heredia Canales}
    \author[kit]{Valentin Kozlov}
    \author[upv]{Sergio Langarita}
    \author[cnca]{João Machado}
    \author[predictia]{Daniel San Martín}
    \author[upv]{Germ\'an Molt\'o}
    \author[stuba]{Giang Nguyen}
    \author[ifca]{Marta Obregón Ruiz}
    \author[pnsc]{Marcin Płóciennik}
    \author[ifca]{Susana Rebolledo Ruiz}
    \author[upv]{Vicente Rodriguez}
    \author[ifca]{Judith Sáinz-Pardo Díaz}
    \author[iisas]{Martin Šeleng}
    \author[iisas]{Viet Tran}

    \fntext[first]{First and second authors contributed equally to this work. Other authors are listed in alphabetical order.}

    \affiliation[ifca]{organization={Instituto de Física de Cantabria (IFCA), CSIC-UC},
    addressline={Avda. los Castros s/n}, city={Santander}, postcode={39006}, state={Cantabria}, country={Spain}}

    \affiliation[upv]{organization={Instituto de Instrumentación para Imagen Molecular (I3M), Centro Mixto CSIC - Universitat Politècnica de València (UPV)},
    addressline={Camino de Vera s/n}, city={Valencia}, postcode={46022}, state={Valencia}, country={Spain}}

    \affiliation[iisas]{organization={Institute of Informatics, Slovak Academy of Sciences (IISAS)},
    addressline={Dúbravská cesta 9}, city={Bratislava}, postcode={84507}, country={Slovakia}}

    \affiliation[stuba]{organization={Faculty of Informatics and Information Technologies, Slovak University of Technology},
    addressline={Ilkovičova 2}, city={Bratislava}, postcode={84216}, country={Slovakia}}

    \affiliation[infn]{organization={Istituto Nazionale di Fisica Nucleare (INFN)},
    addressline={Via Enrico Fermi 40}, city={Frascati}, postcode={00044}, state={Roma}, country={Italy}}

    \affiliation[predictia]{organization={Predictia Intelligent Data Solutions},
    addressline={Fernando de los Ríos 48}, city={Santander}, postcode={39006}, state={Cantabria}, country={Spain}}

    \affiliation[kit]{organization={Karlsruher Institut für Technologie},
    addressline={Kaiserstraße 12}, city={Karlsruhe}, postcode={76131}, country={Germany}}

    \affiliation[lip]{organization={Laboratório de Instrumentação e Física Experimental de Partículas},
    addressline={Av. Prof. Gama Pinto 2}, city={Lisboa}, postcode={1649-003}, country={Portugal}}

    \affiliation[pnsc]{organization={Poznańskie Centrum Superkomputerowo Sieciowe},
    addressline={Jana Pawła II 10}, city={Poznan}, postcode={61-139}, country={Poland}}

    \affiliation[cnca]{organization={Centro Nacional de Computação Avançada (CNCA)},
    addressline={Avenida do Brasil, 101}, city={Lisboa}, postcode={1700-066}, country={Portugal}}

    %% Abstract
    \begin{abstract}
        The rapid growth of Artificial Intelligence and Machine Learning in scientific research has highlighted a gap between industry-standard MLOps tools and platforms, and the unique requirements of modern and Open Science, particularly regarding the FAIR (Findable, Accessible, Interoperable, and Reusable) principles. This paper presents AI4EOSC, a federated, open-source platform designed to operationalize the full AI/ML lifecycle within the European Open Science Cloud (EOSC) ecosystem.
        Our methodology tackles the fragmentation of distributed research infrastructures by integrating a modular and distributed architecture comprising an AI development platform, a serverless AI-as-a-Service layer, and a federated orchestration model that is able to integrate heterogeneous compute and storage resources from distributed e-Infrastructures. AI4EOSC also introduces a ``FAIR-by-design'' approach that enforces metadata standardization (via MLDCAT-AP) and W3C PROV-compliant provenance tracking through a platform-integrated CI/CD pipeline.
        AI4EOSC added value is demonstrated through the delivery of a diverse set of community installations, showing consistent and seamless deployment across heterogeneous cloud providers. These installations are validated by a set of scientific cases, showing how our work reduces the manual burden on researchers while ensuring high levels of reproducibility and interoperability and providing an unified environment for development, training, and production of AI/ML models in the EOSC.
    \end{abstract}

    %  %\includegraphics{grabs}

    %  \item Research highlight 1 \item Research highlight 2

    %% Keywords
\end{frontmatter}

\section{Introduction}
\label{sec:introduction}

The adoption of Artificial Intelligence (AI) and Machine Learning (ML) in scientific research has moved beyond early experimentation. The impact of these techniques over the last years has revolutionized the field in a wide range of applications ---e.g., by making it possible to tackle new scientific endeavors \cite{chervonyi2025gold, jumper2021highly, espeholt2022deep}, by automatizing manual and tedious tasks \cite{KHOGALI2023102232} or by speeding up some time-consuming and effort-intensive tasks \cite{Sergeyuk_2025}, just to cite some examples. These techniques are a reality, and they form nowadays part of the basic set of tools for a wide set of scientists and researchers \cite{hao2024aiexpandsscientistsimpact}. However, this rapid  adoption has exposed a critical gap: the infrastructure and practices needed to develop, share, and deploy AI/ML models in a reproducible, transparent and interoperable manner are still missing, or remain fragmented, within the scientific community.

This gap can be identified across two different axes. On the one hand, there is a large lifecycle fragmentation: the tools that are available for researchers working on AI/ML (integrated development environments, experiment tracking, model repositories, compute infrastructures, etc.) are normally independent components that work in isolation, each of them addressing a different AI/ML lifecycle phase. Assembling all these components together, so that they can support a researcher through the development of their pipelines, poses a significant technological burden on individual scientists. On the other hand, from the scientific standpoint, these tools have been developed without the specific needs of scientific research, and in particular there is a clear absence of \textit{FAIR-by-design} enforcement. The FAIR principles (that stand for Findable, Accessible, Interoperable, Reusable) \cite{wilkinson2016fair} have become a key pillar for open science practices, and nowadays are the norm in scientific research\footnote{The FAIR principles aim to provide guiding principles and standards designed to optimize the reuse of digital assets}. However, in spite of this fact, their application on AI/ML assets remain largely manual and hence its adoption is uneven.

We argue that these two axes cannot be treated as independent problems that can be tackled separately. General data platforms address data-level FAIRness, but they are not well suited to the dynamic nature of model development, where provenance, metadata, and deployment artifacts evolve continuously. FAIR compliance of AI/ML assets cannot be easily achieved by just adding metadata on top of a fragmented pipeline, and structural (i.e., FAIR-by-design) enforcement is needed. Metadata generation and their interoperability is only easily achievable if a single platform controls how assets are generated, built, versioned, and published; provenance records are only meaningful if they can capture the whole chain from data ingestion through training to deployment; interoperability cannot be achieved if there are vendor lock-ins or other trade-offs. This creates a significant architectural dependence, meaning that end-to-end integration is a necessary requirement in order to achieve a FAIR-by-design enforcement. To the best of our knowledge, there is not an open-source platform that addresses both axes together, particularly within the federated, multi-institutional context of European research infrastructures.

Machine Learning Operations (MLOps) is the closest analogue from the industry. MLOps has emerged as a set of practices and tools designed to ease the management of the entire AI/ML lifecycle \cite{berberi2025machine}, analogous to how DevOps bridges the gap between software development and IT operations. While industrial MLOps platforms have been successfully adopted they are are designed for cluster-locked and vendor-controlled environments, lacking native federation across heterogeneous and distributed e-Infrastructures. Moreover, its adoption in the scientific and research context is not that prominent and the systems supporting it prioritize operational performance over the scientific transparency, open standards, FAIR compliance that research demands.

This gap is particularly critical in the context of the European Open Science Cloud (EOSC) ecosystem. The EOSC is an European Commission initiative aimed at creating the ``web of FAIR data and services'', as a transformative process towards the promotion of open science practices among researchers, also delivering a comprehensive environment for the sharing, processing, and analysis of research data \cite{doi/10.2777/935288}. In the context of AI/ML applications and in the EOSC ecosystem, the heterogeneity of compute providers and the diversity of scientific communities makes more complex to reach this ambitious vision, and tackling it from a simple infrastructure standpoint is not enough, requiring of platforms that can operationalize open science principles across the entire AI/ML lifecycle.

In this context, we present the AI4EOSC platform, a federated open-source platform that tackles both axes of the gap simultaneously. The central architectural contribution of this work is the integration of the full AI/ML lifecycle (going from interactive development environments and federated training, through serverless deployment and drift monitoring, to W3C PROV-compliant provenance tracking and semantic interoperability via MLDCAT-AP) into a unified system where FAIR principles are enforced at the infrastructure level, automatically and transparently, without placing additional burden on researchers. AI4EOSC is tightly integrated within the EOSC ecosystem, spans the cloud-to-edge continuum, integrates natively with EOSC research infrastructures, and has been deployed and validated in production by three independent scientific communities.

The remainder of this manuscript is structured as follows:
Section~\ref{sec:related} introduces the background and related work. In
Section~\ref{sec:architecture} and Section~\ref{sec:desgin-considerations}, we discuss the general architecture of the AI4EOSC platform and the design  considerations that have underpinned its development.
In Section~\ref{sec:performance} we provide performance metrics that validate the robustness of the platform.
In  Section~\ref{sec:usage-scenarios} we explain how different scientists can leverage the platform across the full AI/ML lifecycle.
Finally, Section~\ref{sec:conclusion} draws out the main conclusions and future work.

\section{Related work and motivation}
\label{sec:related}

While there are many private and proprietary platforms to deploy AI/ML workloads, few have focused on addressing the specific open science and reproducibility requirements key to scientific research. The current landscape is characterized by framework fragmentation, where existing solutions address specific parts of the AI/ML lifecycle, often in isolation. We argue that there is a clear lack of integrated, open-science-focused solutions that span the entire lifecycle from development to deployment.

\subsection{ML lifecycle platforms and Industrial MLOps}

Standard industrial or enterprise frameworks like Kubeflow \cite{kubeflow} or Polyaxon \cite{polyaxon} provide robust automation but remain largely ``cluster-locked'' to specific Kubernetes environments. Federation of these clusters is not possible without the usage of additional tools or specialized control planes. This restricts their utility for executing workloads across the geographically distributed research e-Infrastructures typically used in scientific collaborations. Furthermore, while some the existing tools offer experiment tracking (either open source like MLflow \cite{zaharia2018mlflow} or LabML \cite{labml}, or proprietary ones like Weights and Biases \cite{wandb}, Valohai \cite{valohai} or Comet \cite{comet}), their focus remains on industrial workflows with limited support for open science and automated FAIR (Findable, Accessible, Interoperable, and Reusable) principles.

Moreover, end-to-end support for the whole AI/ML lifecycle process exists as commercial and closed source products (e.g., H2O.ai \cite{h2O_ai}) and restricts model deployment or the data storage to their specific cloud solutions (like Google Vertex AI \cite{google_vertex}, Amazon Sage Maker \cite{amazon_sagemaker} or Azure ML Studio \cite{azure_ai}), thus defeating the open principles on which science is built. Similarly, model serving solutions like Seldon Core \cite{seldoncore} prioritize operational performance over the scientific transparency and deep tracking required for research trustworthiness.

\subsection{Model Repositories and Scientific Catalogs}

General-purpose repositories like Hugging Face \cite{huggingface}, OpenML \cite{OpenML2025} (an open platform to share datasets and AI/ML models) and specialized scientific catalogs like the BioImage Model Zoo \cite{ouyang2022bioimage} or Kipoi \cite{Avsec2018KipoiAT} have demonstrated as essential for sharing assets with a broader community. However, model metadata is often unstructured and limited and the repositories are often detached from the training process, making it impossible for users to assess the full lineage or provenance of a model. AI4EOSC overcomes this by integrating the repository directly with a platform-enforced CI/CD provenance pipeline, ensuring users can fully assess the provenance of any given asset.

\subsubsection{Provenance in ML Pipelines}

The AI4EOSC provenance system shares its core objective with other PROV-based ML pipeline initiatives, most notably yProv \cite{padovani_provenance_2025}, which also treats provenance as a first-class concern and aims to capture end-to-end relationships between data, models, and processing steps. However, the two approaches reflect a fundamental design trade-off. yProv achieves fine-grained lineage tracking by integrating tightly with the execution environment, instrumenting pipeline components to emit provenance information at runtime. This enables detailed capture of model evolution, parameter tuning, and intermediate data transformations, making it well suited to controlled settings where the full pipeline is under unified management.

AI4EOSC takes the opposite approach: provenance is collected non-intrusively from metadata already exposed by existing platform components, without requiring any modification to their internal behavior. While this constrains the level of detail to what those components natively provide, it offers significantly greater flexibility and ease of adoption in heterogeneous, real-world deployment environments where pipeline components are diverse and independently operated. Furthermore, the AI4EOSC approach explicitly targets semantic interoperability by producing PROV-O compliant JSON-LD output and supporting alignment with domain-specific ontologies, enabling provenance information to be reused and queried across system boundaries rather than remaining confined to the originating pipeline.

\subsection{Evolution from the DEEP Platform}

The DEEP-Hybrid-DataCloud (DEEP) \cite{garcia2020cloud} platform pioneered a distributed architecture for transparent access to European e-Infrastructures with focus on scientific applications. While DEEP successfully provided standardized APIs (DEEPaaS) and eased access to accelerators (GPUs), it lacked the deep semantic interoperability and automated metadata generation now required within the EOSC federation and ecosystem. AI4EOSC represents a significant evolution, introducing a serverless AI-as-a-Service (AIaaS) layer, federated learning (FL) capabilities, and a comprehensive FAIR-by-design framework that was previously absent.

\subsection{Open Science and FAIR Principles}

The \emph{FAIR principles} have become a central pillar of modern open-science practices. In the context of AI/ML research, these aspects should not be considered just as a mere set of best practices, but as an important vehicle towards AI/ML model transparency, reproducibility and trustworthiness. While general data platforms like Zenodo or Dataverse address data-related FAIRness (such as data provenance or licensing) \cite{esfri_eosc_task_force_2024_10980285}, they are often unable to respond to the dynamic nature of AI/ML model development and deployment, hence limiting the FAIR principles adoption beyond data. Researchers that want to ensure the FAIRness of their complete AI/ML solutions must bundle together a set of tools, standards and practices, at the cost of reproducibility, efficiency of their research, or compliance with open science principles. AI4EOSC addresses this by operationalizing these principles across the entire AI/ML lifecycle through automated, ``FAIR-by-design'' mechanisms, that are easily extensible in order to accommodate new or unforeseen requirements.

\subsection{AI4EOSC motivation}

In this fragmented context, AI4EOSC addresses the need for an approachable, open-source solution that streamlines model creation for both technical and non-technical scientists. The platform differentiates itself through several core architectural innovations (as described in Section~\ref{sec:architecture}):

\begin{itemize}
    \item \textbf{Transparent Inter-Cluster Federation:} Our federated workload management system is able to scale across many physically distributed datacenters and e-Infrastructures, with a variety of deployment and execution methods, via a single and unified control plane that avoids the infrastructure silos of standard MLOps stacks.

    \item \textbf{Cloud-to-Edge Continuum:} AI4EOSC mobilizes resources across a continuum, enabling deployment on heterogeneous hardware including low-capacity edge devices and cloud computing providers.

    \item \textbf{Hardened Federated Learning support:} The platform extends standard Federated Learning (FL) with token-based authentication, novel aggregation methods, divergence monitoring and server-side differential privacy and metric privacy notions to defend against client inference attacks.

    \item \textbf{Integrated Provenance Tracking:} The platform automatically transforms training fragments into W3C PROV-compliant RDF graphs, providing lineage traceability that is currently detached or proprietary in industry platforms.

    \item \textbf{Semantic Interoperability:} AI4EOSC leverages MLDCAT-AP in order to deliver interoperability at the semantic level with other platforms and data spaces, enabling for seamless interactions with other marketplaces.

    \item \textbf{Integrated Drift Monitoring:} Through the \textit{Frouros} library and the \textit{DriftWatch} system, researchers can detect and visualize concept and data drift at inference time.

    \item \textbf{Seamless EOSC Integration:} AI4EOSC is natively integrated with the EOSC, enabling researchers to leverage EOSC's infrastructure for data sharing (e.g., EOSC EU Node resources), collaboration, or long-term preservation. This integration ensures that AI/ML research is not only FAIR but also interoperable with the broader European research ecosystem.
\end{itemize}

To synthesize these differences and explicitly contrast AI4EOSC with the state-of-the-art, Table~\ref{tab:comparison_comprehensive} provides a feature-wise comparison across key technical and scientific dimensions.

\begin{table*}[h]
\centering
\resizebox{\linewidth}{!}{
\begin{tabular}{lcccccccc}
\toprule
 & \textbf{Kubeflow} & \textbf{Polyaxon} & \textbf{Seldon} & \textbf{SageMaker} & \textbf{OpenML} & \textbf{Hugging} & \textbf{DEEP} & \textbf{AI4EOSC} \\
 \textbf{\textit{Feature}} & \cite{kubeflow} & \cite{polyaxon} & \cite{seldoncore} & \cite{amazon_sagemaker} & \cite{OpenML2025} & \textbf{Face}  \cite{huggingface} & \cite{garcia2020cloud} & \textbf{(This work)} \\
\midrule
\textbf{\textit{Scientific Focus}} & Low & Low & Low & Low & High & Medium & High & \textbf{High} \\
\textbf{\textit{Openness}} & Full & Open-core & Open-core  & Closed & Full & Partial & Full & \textbf{Full} \\
\textbf{\textit{FAIR-by-Design}} & No & No & No & No & Partial & No & Low & \textbf{Native} \\
\textbf{\textit{Provenance}} & Limited & Limited & Limited & Proprietary & Limited & Limited & Low & \textbf{W3C PROV} \\
\textbf{\textit{LLM Integration}} & Partial & Partial & Partial & High & No & High & No & \textbf{Native (RAG)} \\
\textbf{\textit{Learning Schemes}} & Dist. & Dist. & N/A & Dist. & N/A & N/A & Dist. & \textbf{Fed. \& Dist.} \\
\textbf{\textit{Drift Detection}} & External & Limited & Yes & Yes & No & No & No & \textbf{Native} \\
\textbf{\textit{Infrastructure}} & Kubernetes & Kubernetes & K8s/Cloud & Prop. Cloud & N/A & N/A & e-Infras & \textbf{e-Infras/Edge} \\
\textbf{\textit{Federation}} & Cluster-lock & Cluster-lock & Cluster-lock & Vendor-lock & N/A & N/A & Partial & \textbf{Inter-cluster} \\
\textbf{\textit{Delivery Model}} & Software & Both & Both & Platform & Both & Platform & Both & \textbf{Both} \\
\textbf{\textit{EOSC integration}} & No & No & No & No & No & No & Partial & \textbf{Native} \\
\bottomrule
\end{tabular}}
\caption{Feature-wise comparison of AI4EOSC against representative industrial MLOps stacks, scientific tools, and the predecessor DEEP platform.}
\label{tab:comparison_comprehensive}
\end{table*}

\section{The AI4EOSC platform architecture}
\label{sec:architecture}

In the following section, we describe the architecture of AI4EOSC using a simplified view of the C4 Model \cite{brown_c4_nodate} to offer a clear and structured visualization of its components and interactions. This simplified representation aims to enhance the article readability. For a detailed and comprehensive view of the AI4EOSC architecture, please refer to the complete C4 implementation available at \cite{alvaro_lopez_garcia_2024_13343448}.

\begin{figure*}
    \centering
    \includegraphics[width=0.8\linewidth]{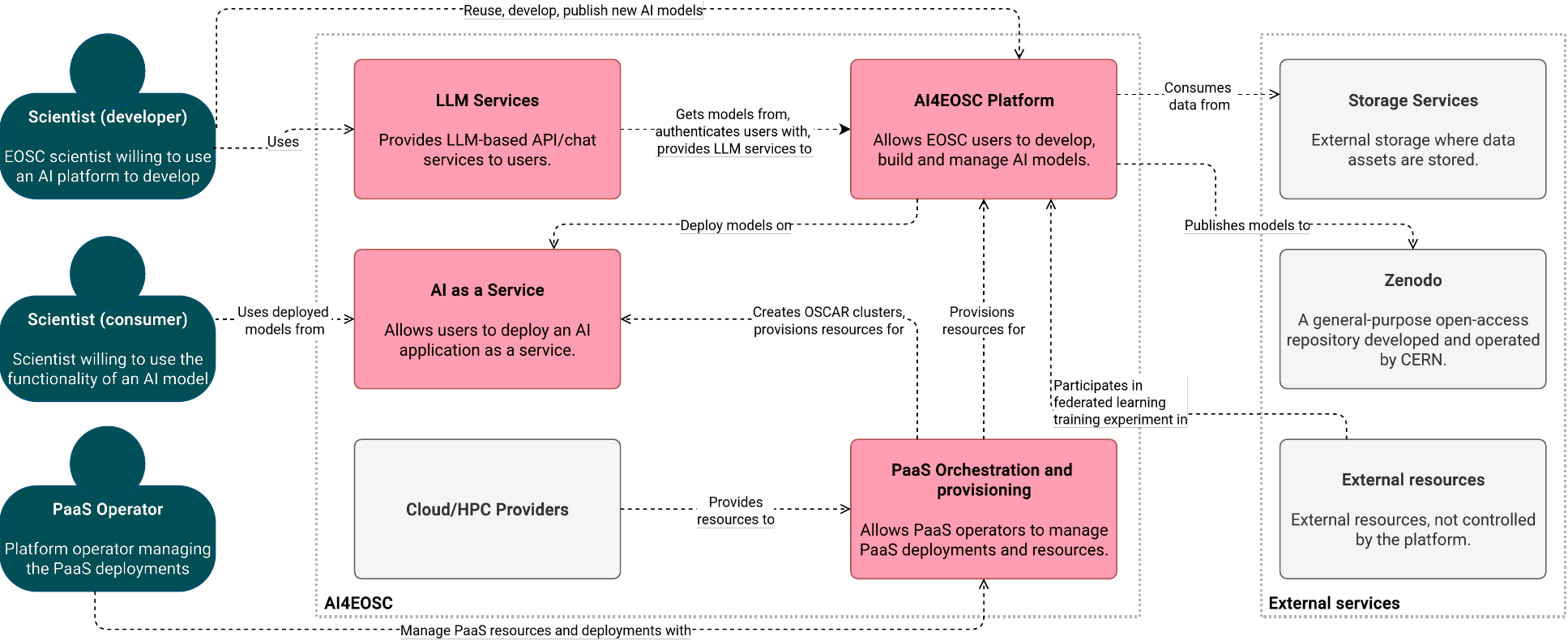}
    \caption{Simplified view of the AI4EOSC architecture system landscape view (C4 notation).}
    \label{fig:arch:system-view}
\end{figure*}

\begin{figure}
    \centering
    \includegraphics[width=0.9\linewidth]{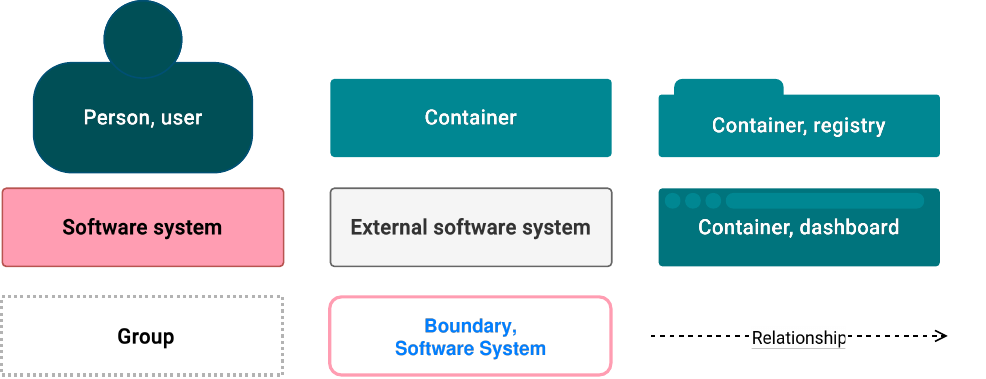}
    \caption{Legend for all C4 diagrams used in this manuscript.}
    \label{fig:legend}
\end{figure}

Our architecture follows a modular design, organized into four distinct systems: AI4EOSC Development Platform, AI4EOSC AI/ML as a Service, AI4EOSC LLM services, and AI4EOSC Platform Orchestration. These components work interconnected in order to be able to deliver an specific functionality, but being integrated with the rest of the systems. This modularity is illustrated in Figure~\ref{fig:arch:system-view}. The legend for all the C4 diagrams is shown in Figure~\ref{fig:legend}.

\subsection{AI4EOSC Development Platform system}

\begin{figure*}
    \centering
    \includegraphics[width=1\linewidth]{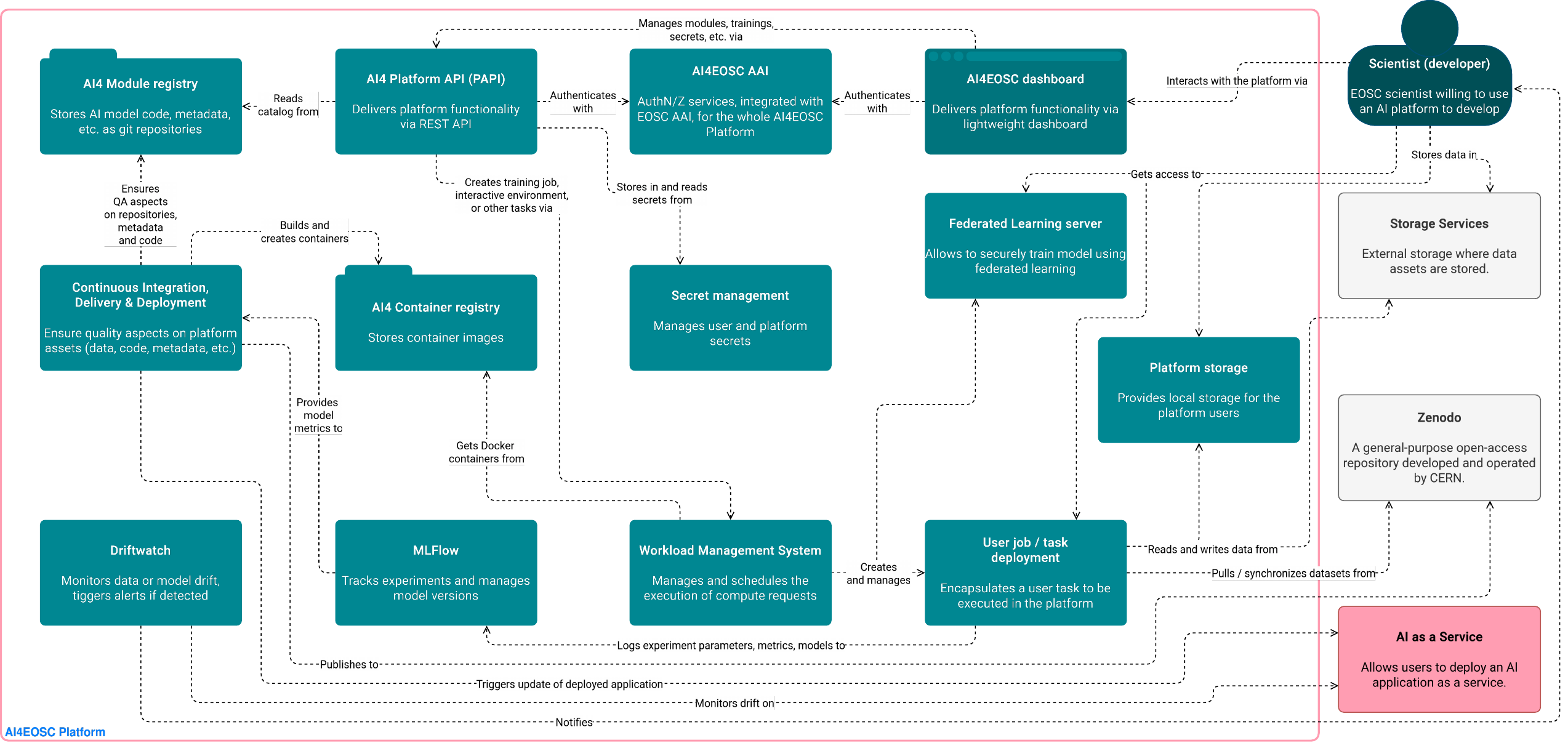}
    \caption{Simplified AI4EOSC Development platform architecture: container view (C4 notation).}
    \label{fig:ai4eosc}
\end{figure*}

This system, depicted in Figure~\ref{fig:ai4eosc}, includes all the needed services to deliver a comprehensive platform and toolbox to enable researchers to build, develop, train and share AI/ML models following the FAIR principles. It also includes some core, common and shared services that conform the main AI4EOSC control plane, that serve as auxiliary services to both this system and others. The most relevant components are explained next subsections.

\subsubsection{Workload Management System}
\label{sec:workload-management}

The Workload Management System (WMS) is a federation of service providers that is used to manage the platform resources and execute the platform jobs. Given the federated nature of the platform, we have chosen a set of components that allow us to federate geographically distributed resources easily in the form of a service mesh. Therefore, in order to address the challenges and complexity of cross-datacenter resource federation, the AI4EOSC architecture leverages a decentralized WMS based on Hashicorp Nomad and Consul. Unlike centralized orchestrators, like Kubernetes, this combination allows for a lightweight scheduling across geographically distributed providers with minimal administrative overhead. Moreover, both the inter-cluster orchestration and intra-cluster scheduling is managed from the same control plane, without the need of additional layers or components. The current federation includes resources from over 5 different clouds located in 5 different countries: IFCA-CSIC (Spain), IISAS (Slovakia), INCD (Portugal), Tubitak (Turkey) and PNSC (Poland). On top of Nomad node failure monitoring, the AI4EOSC platform runs periodic quality checks (including configuration and performance) to remove low performing nodes from the federation.

The AI4EOSC WMS handles two primary types of executions: user-oriented workloads and system-level management tasks. User workloads refer to both AI modules (\ref{sec:ai-modules}) and other specialized tools (\ref{sec:tools}) that the platform offers to enrich the user experince. Both of these can be executed in two distinct modes. On the one hand, in \textit{standard} mode resources are persistent, providing researchers with continuous access via Interactive Development Environments (IDEs) or APIs. On the other hand, the \textit{batch} mode utilizes a transient allocation strategy, where resources are provisioned strictly for the duration of a training process to maximize global cluster efficiency and resource sharing.

Besides, in order to enhance user experience and provide structural functionality to the platform, AI4EOSC augments Nomad jobs with transparent sidecar tasks. These components manage additional function that is not directly exposed the users, such as automated connectivity to external storage systems, perform dataset synchronization from remote repositories at runtime, and trigger user notifications. Moreover, system-level jobs including Traefik reverse proxies, the Platform API, and the administrative Dashboard (\ref{sec:papi}) are also part of  these workloads to maintain platform stability and service accessibility without direct user intervention.

\subsubsection{Platform API and dashboard}
\label{sec:papi}\label{sec:dashboard}

The Platform API (PAPI) is the component that delivers the platform functionality via a well-defined set of endpoints, following a REST architecture. It is developed following the OpenAPI specifications, providing the formal description and documentation of the PAPI functionality.

Built on top of the API, the AI4EOSC Dashboard is the main user entry-point to the AI4EOSC Platform. Both PAPI and the dashboard are built with extensibility in mind, in order to be able to easily integrate new functionality without interfering with the current one.

\begin{figure}
    \centering
    \includegraphics[width=1\linewidth]{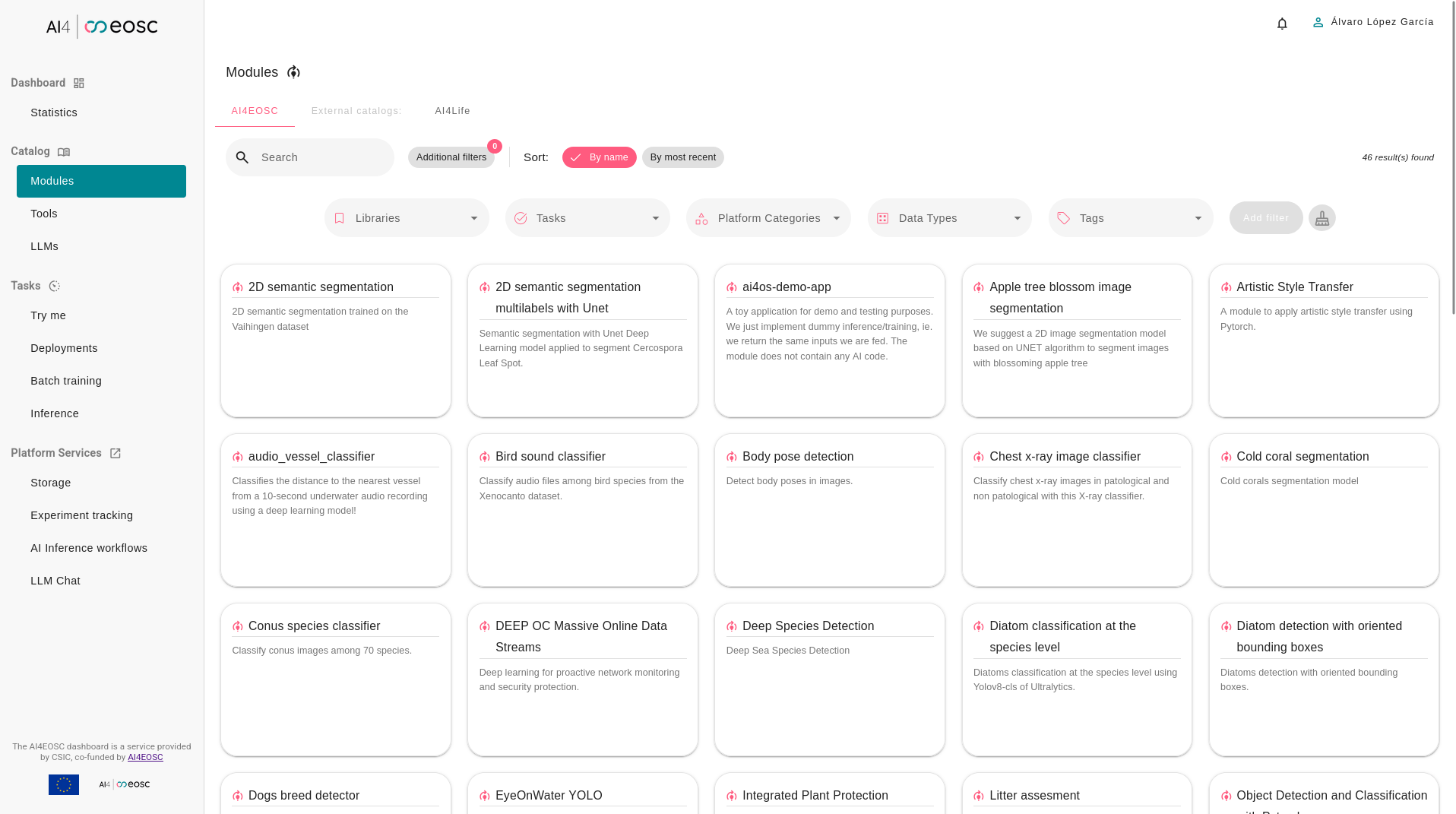}
    \caption{A snapshot of the AI4EOSC dashboard (available at \url{https://dashboard.cloud.ai4eosc.eu}) providing an overview of the existing AI modules in the platform.}
    \label{fig:dashboard}
\end{figure}

The Dashboard (Figure~\ref{fig:dashboard}) allows users to easily explore and view the associated metadata for the different catalogs available in the system: AI modules (\ref{sec:ai-modules}), tools (\ref{sec:tools}), LLM services (\ref{sec:llm}) and other external AI catalogs (\ref{sec:external-ai-catalogs}). It also provides the ability to launch development environments or codespaces in the platform, batch executions, easy management of existing deployments, platform-wide secret management (e.g., API keys, storage integration) and resource consumption reporting, among other user-facing functionalities.

\subsubsection{AI Modules}
\label{sec:ai-modules}

The AI modules are the main platform asset that users interact with, as they encapsulate the AI/ML models that the platform is able to manage. They are both produced and consumed by the users, although the AI4EOSC development team also maintains reference implementations of relevant AI models. They are created and stored as git repositories (currently hosted on GitHub, but any other software forge can be used) allowing users to easily manage and upload new updates to the modules, as well as making it possible to enforce quality assurance mechanisms and MLOps practices (\ref{sec:cicd}) from the platform point of view. To support the development of new AI modules that comply with the AI4EOSC platform and best practices (e.g., metadata standards (\ref{sec:ai4os-metadata})), a set of pre-defined software templates is provided to users.

\subsubsection{Tools}
\label{sec:tools}

Tools are a collection of additional assets that the AI4EOSC Platform provides to users in order to support them during the whole ML development lifecycle. These tools profit from the platform extensibility (\ref{sec:extensibility}) as a means to enhance and augment the current functionalities.

Unlike AI modules (\ref{sec:ai-modules}), tools are developed and curated by the developers of the AI4EOSC Platform, in order to ensure that the integration is implemented correctly. Once integrated, tools are easily self-deployable by users on their personal workspaces, just like AI modules.

For technical reasons, some of these tools do not make sense to be deployed per-user and therefore are instead offered as platform-wide services, like the experiment tracking service or the drift monitoring (\ref{sec:drift-detection}); however the integration mechanism remains the same.

In the following, we will explain the main set of tools offered by the platform.

\paragraph{Development Environment -- Codespaces}
\label{sec:development-environment}

The AI4EOSC Platform allows users to deploy a ready-to-use development environment, referred to as Codespaces, to create their new AI modules (\ref{sec:ai-modules}). The environment can be used out-of-the-box with multiple versions of the main Deep Learning frameworks such as TensorFlow \cite{abadi2016tensorflow},
PyTorch \cite{2019pytorch}. Additional frameworks are also supported by starting from provided plain Ubuntu or NVIDIA CUDA flavors, or by installing a custom software stack.

When configuring the deployment, users can choose their preferred IDE for code development. The platform currently supports both Visual Studio Code (VSCode) bundled with many useful extensions, and installed using Code Server, and JupyterLab. All IDE endpoints are password or token protected to prevent malicious actors from accessing user data and code.

\paragraph{Federated Learning servers}
\label{sec:federated-learning-servers}

The AI4EOSC Platform allows executing federated learning servers \cite{nguyen2025landscape}, allowing AI/ML models to be trained collaboratively across multiple institutions without sharing private data. The platform currently supports the two most widely used and robust frameworks: Flower \cite{beutel2020flower} and NVFLARE \cite{roth2022nvidia}. Both FL servers can be launched interactively from the dashboard, and FL clients can be deployed within the AI4EOSC Platform or connect from outside, thus easily accommodating the individual privacy needs of the different data owners. This flexible and distributed approach allows leveraging a wide range of resources in the Cloud continuum that could not support a full ML training cycle by themselves due to their low capacity (e.g., edge devices).

Beyond the standard framework capabilities, AI4EOSC has extended the Flower framework with a set of platform-level enhancements. First, we have implemented token-based client-server authentication \cite{sainz-pardo_diaz_making_2024}, managed directly from the dashboard via the platform secret management system. This allows operators to create and revoke client tokens at any time, ensuring that only authorized clients can participate in the training. Second, the server incorporates a mechanism to monitor the divergence between the model weights of different clients, allowing the detection of clients that deviate significantly from the rest and may be introducing poisoning behavior, with the option to revoke their access token immediately. Third, server-side differential privacy can be applied during aggregation using fixed-clipping Gaussian mechanisms, and the platform further incorporates the notion of server-side metric differential privacy \cite{SAINZPARDODIAZ2026115993}, which scales the noise multiplier dynamically based on a distance metric computed from local model parameters, protecting against client inference attacks. Finally, AI4EOSC introduces \textit{FedAvgOpt}
\cite{diaz2025enhancingconvergencefederatedlearning}, a novel aggregation method that optimizes the aggregated weights to be as close as possible to the individual client weights using the Nelder-Mead method, improving convergence without compromising privacy. The supported aggregation strategies (i.e., \textit{FedAvg}, \textit{FedAvgM}, \textit{FedMedian}, \textit{FedProx}, \textit{FedOpt}, \textit{FedAdam}, \textit{FedYogi}, and \textit{FedAvgOpt}) are selectable directly from the dashboard.

The security and privacy implications of this learning approach, including its role in defending against poisoning and inference attacks, are further discussed in Section~\ref{sec:privacy_fl}.

\subsubsection{MLOps practices}

Extending the DevOps approach to the ML realm, AI4EOSC delivers a series of tools to allow the users to address the key challenges that specifically arise during the ML cycle.

\paragraph{Continuous Integration and Delivery of models}
\label{sec:cicd}

A key component of the AI4EOSC platform, that underpins the MLOps model management is the CI/CD system comprising both platform-enforced and user defined pipelines executed via Jenkins. This system ensures that any AI module integrated in the Marketplace satisfies the same consistent requirements: e.g. by ensuring that published models include the needed metadata, or triggering updates on existing services based on a given module.

To achieve this, AI4EOSC maintains a platform-wide pipeline definition that is enforced in all new and existing modules. This pipeline is executed only after the user-defined jobs are executed, and it is optimized in order to execute different tests according to the changes that trigger them, such as metadata updates only or the modification to the source code.

\begin{figure}[t]
    \centering
    \includegraphics[width=1\linewidth]{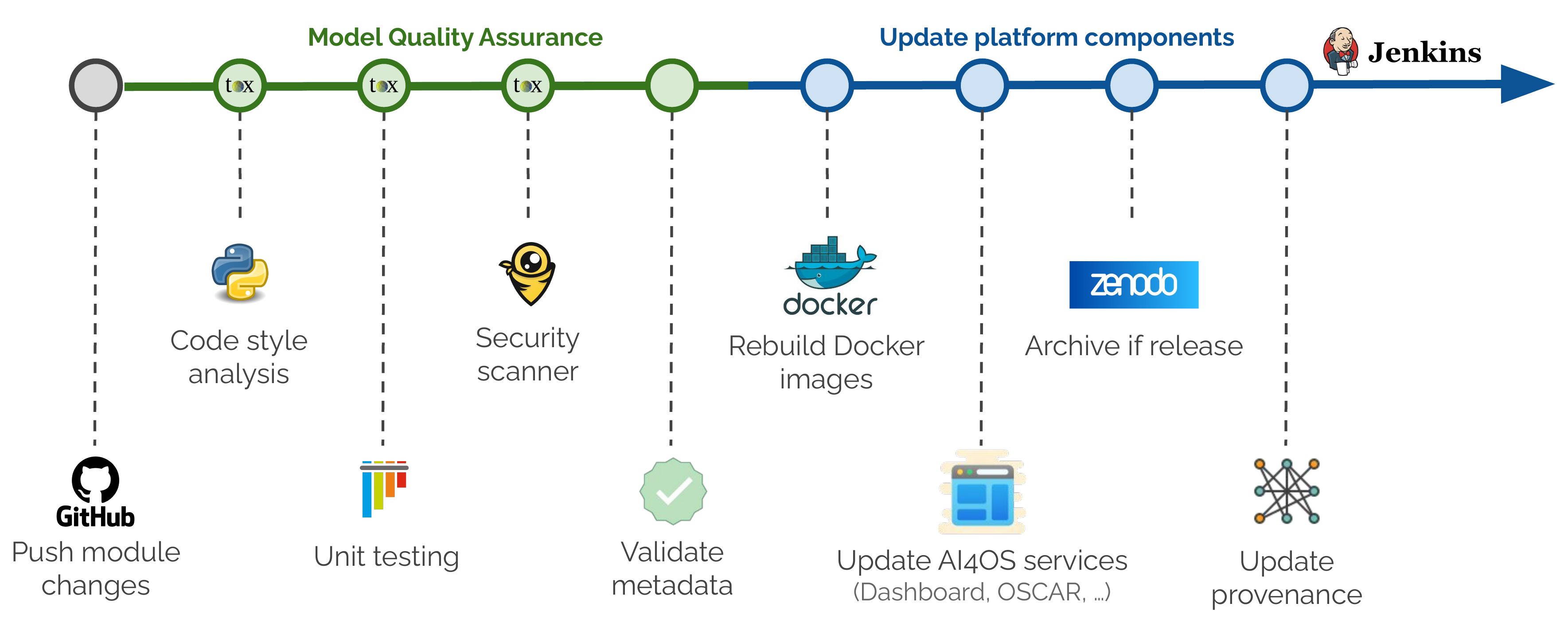}
    \caption{CI/CD pipeline for AI modules}
    \label{fig:jenkins-pipeline}
\end{figure}

The platform-level CI/CD pipeline is illustrated in Figure \ref{fig:jenkins-pipeline}. As shown, there are a set of steps that comprise software and model quality assurance (e.g., code style, unit testing, security scanning, metadata validation), and a second set of steps that are focused on the delivery of the model in the corresponding components, including keeping track of the steps in the provenance system. From a platform perspective, it ensures the following key aspects:
\begin{itemize}
    \item Metadata validation: We ensure that the user-defined metadata (\ref{sec:ai4os-metadata}) is compliant with the metadata schema that the platform relies on. This is a critical step, taking into account that the PAPI, Dashboard, and the interoperability with other metadata schemes require that accurate and consistent metadata is in place.

    \item Docker images generation and publication: The platform relies on Docker containers to ensure that the modules are self-contained and executable in a wide variety of systems. Therefore we trigger the build of new images and their publication in the platform Harbor container registry\footnote{\url{https://registry.cloud.ai4eosc.eu/}} and Docker Hub\footnote{\url{https://hub.docker.com/u/ai4oshub}}. To avoid unnecessary builds, the rebuild step is only triggered when non-metadata changes are detected and can be disabled in specific git branches.

    \item Zenodo release integration: In order to ensure that the code remains accessible, and following the FAIR principles also for software and model assets, we ensure that the Zenodo-GitHub integration is in place. If it is not, we create a deposit and directly upload any pending releases.

    \item Notification to other platform services: As we will explain in Section~\ref{sec:ai-service-pipelines}, the platform provides an AI as a Service component that allows to deliver the AI/ML models functionality following a serverless model. In order to let them know that a new version of the model is available, we notify the services via defined web-hooks, providing an updated metadata version, so that they can react to these changes and update the running modules if they are configured to do so. Moreover, we also notify the PAPI, in order to trigger an update of the cached model metadata.

    \item Notification of provenance updates: The AI4EOSC platform keeps track of the model provenance (\ref{sec:provenance}). Therefore, whenever a model is updated, we notify the provenance system, including the CI/CD gathered metadata (e.g., tests executed), ensuring that the provenance graph always reflects the latest changes.
\end{itemize}

\paragraph{Experiment tracking}
\label{sec:experiment-tracking}

Users deploying AI modules (\ref{sec:ai-modules}) or Development Environments (\ref{sec:development-environment}) can log their training metrics into a dedicated MLflow \cite{zaharia2018mlflow} tracking server instance we have deployed, featuring a custom Authentication GUI to be able to integrate it with our authentication system. MLflow credentials are saved in the platform secrets manager and automatically injected in the WMS deployments (\ref{sec:workload-management}) created from the Dashboard (\ref{sec:dashboard}).

By logging multiple training runs and saving the corresponding model weights with proper versioning, we can enhance the reproducibility of the models. This information is later included in the provenance chain, as we will explain in Section~\ref{sec:provenance}.

\paragraph{Drift monitoring}
\label{sec:drift-detection}

AI4EOSC has developed a drift monitoring system, DriftWatch\footnote{\url{https://drift-watch.cloud.ai4eosc.eu/}}. This component allows to monitor AI modules (\ref{sec:ai-modules}) in production, warning the model developer when performance starts to degrade and a model retraining is needed. In order to detect if a drift is happening or not, a detector is needed. Driftwatch is agnostic to the underlying drift detection framework chosen by the developer such as Alibi-detect \cite{Van_Looveren_Alibi_Detect_Algorithms_2024}, Evidently, etc. However, AI4EOSC has developed Frouros \cite{sisniega2024frouros}, a Python library for drift detection in machine learning systems that supports a large combination of classical and more recent algorithms for both concept and data drift detection. When an AI module is developed,
users can define reference clean/faulty data and train a Frouros detector on them (users handling image data would need to first train an autoencoder to reduce the dimensionality of their data). Then, at inference time, when the models are deployed in the production AIaaS (\ref{sec:ai-service-pipelines}), the inference data is passed trough the Frouros detector and the resulting p-values for detection are sent to the DriftWatch server \cite{Cespedes2024drift}. From an intuitive user interface, users are able to log in and visualize whether drift has occurred, eventually inspecting the data that caused the drift.

\subsection{AI/ML as a Service system}
\label{sec:ai-service-pipelines}

\begin{figure}
    \centering
    \includegraphics[width=1\linewidth]{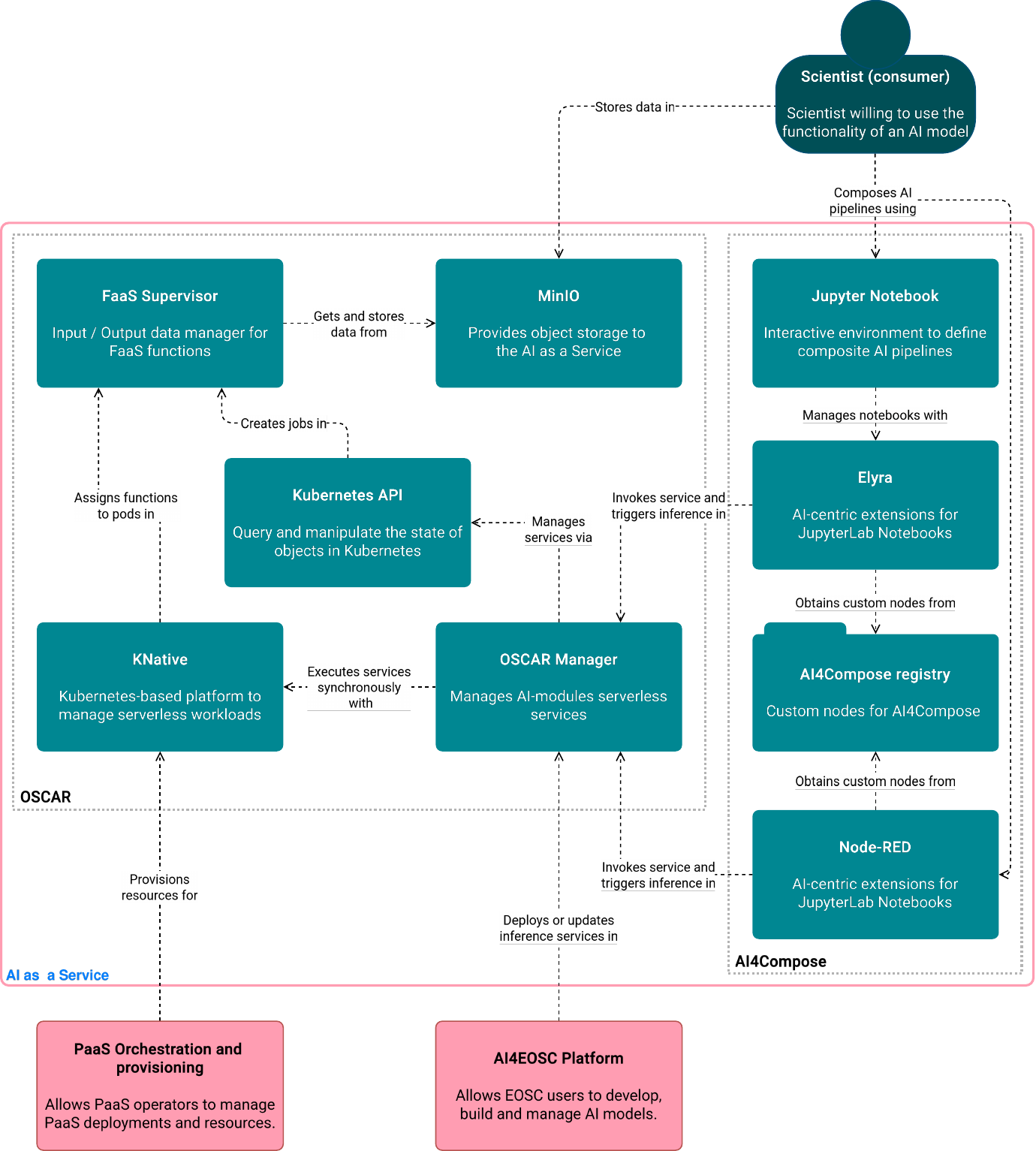}
    \caption{Simplified AI4EOSC AI/ML as a Service architecture: container view (C4 notation).}
    \label{fig:ai4eosc-aiaas}
\end{figure}

The AIaaS, shown in Figure~\ref{fig:ai4eosc-aiaas}, allows user to deploy AI modules (\ref{sec:ai-modules}) in a serverless inference platform, based on OSCAR \cite{perez2019premises}.

The OSCAR cluster allows to autoscale resources on demand, automatically deploying multiple instances of the model to run predictions as needed. The platform allows both synchronous and asynchronous calls to the endpoints via file uploads to a MinIO object store.
The AI4EOSC Dashboard (\ref{sec:dashboard}) allows users to automatically create and manage those serverless endpoints for any AI module in the catalog. Thanks to the DEEPaaS API \cite{garcia2019deepaas} that interfaces the module, users are able to send inference requests to any AI module in a standardized way.

Leveraging on those serverless endpoints, the AI4EOSC Platform allows to create AI pipelines where users can compose multiple inference requests to different AI models, as well as data preprocessing and postprocessing steps.

Pipelines can currently be created with two different visual composition tools: Node-Red, whose instances are managed via the OSCAR cluster, and Elyra, integrated with Jupyter-based environments like the AI4EOSC Development Environment (\ref{sec:development-environment}) or the EGI Notebooks. Users can choose between these two tools to compose their pipelines based on their working environment needs and prior experience.

\subsection{Large Language Models Services system}
\label{sec:llm}

\begin{figure}
    \centering
    \includegraphics[width=\linewidth]{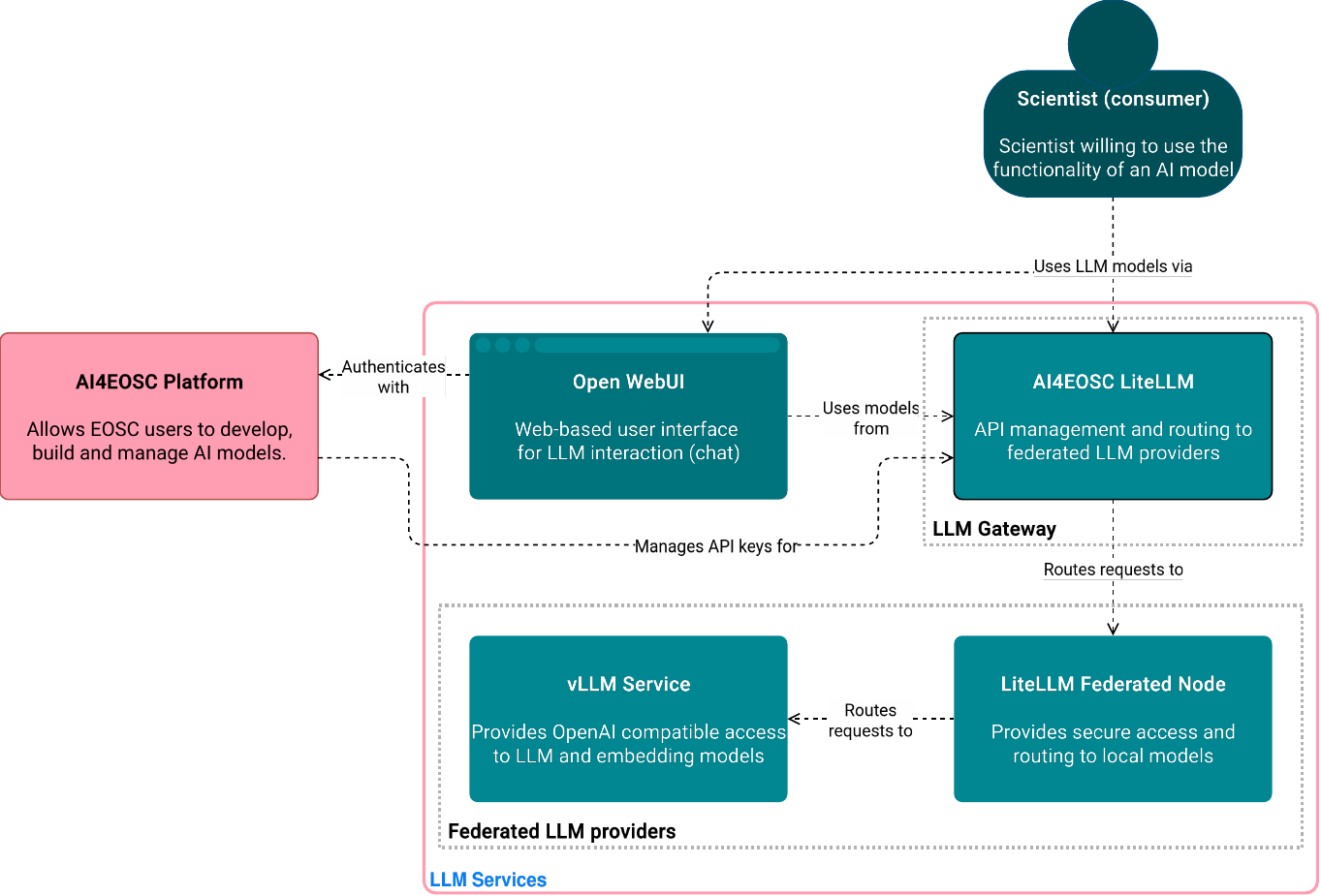}
    \caption{Simplified AI4EOSC Large Language Model Services architecture: container view (C4 notation).}
    \label{fig:ai4eosc-llm}
\end{figure}

This system, described in Figure~\ref{fig:ai4eosc-llm}, is designed to provide a federated inference system focused on the needs of Large Language Models (LLMs). The system leverages vLLM \cite{kwon2023efficient} and LiteLLM in order to provide an easy way to include additional LLM providers, that can be geographically distributed. To achieve this, each resource provider deploys their own LLMs with vLLM and aggregates them with a local LiteLLM instance. Then, in turn, the platform's LiteLLM instance aggregates the different local LiteLLMs, effectively serving all provider LLMs under a single endpoint.

Moreover, the AI4EOSC Dashboard allows users to easily manage API keys to access the service. This system can be directly used via its OpenAI compatible API endpoints, in order to be integrated with any user tool, and via the platform's conversational engine based on OpenWebUI. Moreover, this system is key to provide AI4EOSC platform users with an specific support agent, able to support users with direct answers from the documentation knowledge base.

\subsection{Platform orchestration system}
\label{sec:paas-orchestration}

\begin{figure}
    \centering
    \includegraphics[width=1\linewidth]{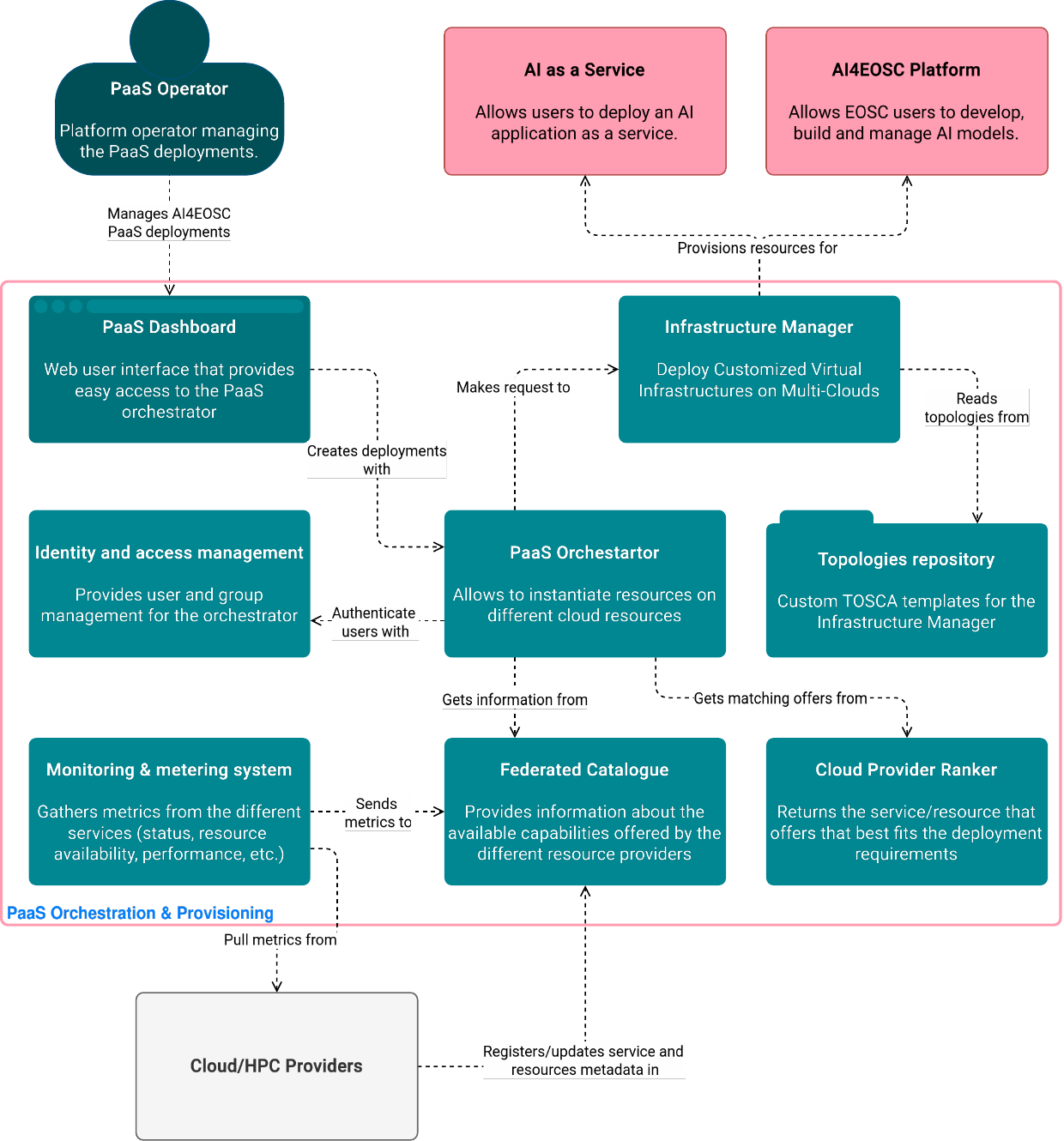}
    \caption{Simplified AI4EOSC platform orchestration architecture: container view (C4 notation).}
    \label{fig:ai4eosc-paas}
\end{figure}

This system, depicted in Figure~\ref{fig:ai4eosc-paas}, is in charge of assisting resource providers and platform operators by automatically provisioning the resources needed to deploy the rest of the platform services. In the AI4EOSC platform, we use the INDIGO-PaaS Orchestration system \cite{costantini2021, salomoni2018indigo} to handle the deployment of platform components across the different IaaS platforms made available by the federated Cloud providers.

The system is composed of different micro-services:

\begin{itemize}
    \item The Infrastructure Manager \cite{caballer2015dynamic,caballer2023infrastructure}, who is responsible for the submission of the deployment to the proper Cloud provider. The provider is selected according to the AI-Ranker \cite{giommi2025}, which generates a ranked list of providers by combining the prediction of deployment success/failure with the estimated deployment creation time.

    \item The Federation-registry \cite{savarese2024}, which is the configuration management database for dynamic cloud federation. Built using FastAPI and neo4j, the Federation-registry is able to efficiently manage the federation-related information and SLA agreements.

    \item The INDIGO-PaaS Orchestrator Dashboard, which provides a user-friendly interface for managing and monitoring the TOSCA-based deployments, built on Flask and MySQL.
\end{itemize}

With the developed TOSCA \cite{tosca} templates, the PaaS Orchestration layer is able to automate the deployment of the platform components, including the Workload Management System (\ref{sec:workload-management}) and the AI as a Service platform (\ref{sec:ai-service-pipelines}).

These strong reproducible recipes make the AI4EOSC platform a suitable solution to be easily adopted by external communities wishing to profit from the technologies developed here. Moreover, the adoption of well known technology standards allow the Cloud-enabled resource providers to easily join the federation and contribute to the existing AI4EOSC platform.

\section{Platform design considerations}
\label{sec:desgin-considerations}

\subsection{Platform FAIR design}

As mentioned when presenting the motivation, the FAIR principles are central to the design of the AI4EOSC platform. Ensuring that AI modules and their associated assets are transparently discoverable, well-described, and reusable by the broader scientific community is a core requirement for a research-oriented platform. AI4EOSC addresses this through standardized metadata schemas, comprehensive end-to-end provenance tracking, and a commitment to openness. This approach ensures that every asset produced by the platform can be understood, reproduced, and built upon by others.

\subsubsection{AI4EOSC metadata}
\label{sec:ai4os-metadata}

In the current data-driven ecosystem, rich metadata is of paramount importance in order to comprehensively describe a given asset in an homogeneous way. Well defined metadata allows also to work towards interoperability, both at the platform and at the AI/ML model level.
To ensure that all the AI modules developed in the platform  provide consistent metadata, the AI4EOSC Platform enforces that each AI module (\ref{sec:ai-modules}) or tool (\ref{sec:tools}) that is registered in the AI4EOSC catalog must include a valid metadata file following a defined schema implemented via JSON Schema.

The AI4EOSC metadata schema includes both user-defined fields (some of them optional) as well as fields automatically filled from external sources like License and modification dates from GitHub. User defined-fields include title, summary, description, DOI, external links (e.g. to dataset or model weights) and tags such as the software libraries used or the data types used as input.

Once integrated, the Platform API retrieves the module's metadata and serializes it in different formats (JSON-LD, RDF Turtle), being able also to transform it into different application profiles, such as the MLDCAT Application Profile. This allows for a greater interoperability of our models with external catalogs, allowing semantic interoperability with formats that are compatible with MLDCAT-AP.

\subsubsection{Provenance}
\label{sec:provenance}

Provenance is essential for science, and a FAIR-by-design system should take into account asset provenance as one of its pillars. AI4EOSC provides an integrated and extensive provenance system that allows transparent traceability of any AI module (\ref{sec:ai-modules}), thus improving their  reproducibility.

The system is designed to operate in a non-intrusive manner, leveraging metadata already generated by existing architectural components such as MLflow and Jenkins, without requiring modifications to these systems. Once a module has committed changes to the Git repository, the CI/CD pipeline triggers the collection of provenance metadata from multiple sources: module metadata (\ref{sec:ai4os-metadata}) and training resources (\ref{sec:workload-management}) from the Platform API  (\ref{sec:papi}), experiment tracking metrics from MLflow (\ref{sec:experiment-tracking}), and build information from the CI/CD pipeline itself (\ref{sec:cicd}).

These heterogeneous, non-semantic metadata fragments are stored in their native JSON representation in a PostgreSQL database. This design choice is deliberate: rather than imposing a rigid schema at ingestion time, the system preserves the original structure of each source, enabling flexible querying and transformation. Semantic interoperability is then achieved by applying RDF Mapping Language (RML) rules via CARML over the stored JSON fragments. These declarative mappings define how disparate metadata elements correspond to a unified semantic model, enabling the integration and linkage of otherwise disconnected provenance information into a coherent end-to-end provenance graph.

The resulting provenance graph is serialized as an RDF document encoded in JSON-LD, ensuring compliance with the PROV-O ontology for standardized provenance modeling. The system also supports alignment with domain-specific ontologies, allowing for contextual enrichment and adaptability across different AI application domains. Having the provenance data in RDF further allows conversion to external formats such as FAIR4ML \cite{fair4ml} or any additional mapping, and can be extended with virtually any information that the platform or the user is able to provide.

A comparative analysis of this design against other PROV-based ML pipeline approaches is presented in Section~\ref{sec:related}.

In order to present the provenance information to users, an interactive web-based visualization tool has been developed and integrated into the AI4EOSC dashboard, providing dynamic, user-friendly interfaces for navigating complex provenance graphs. Additionally, users can query the provenance graph in natural language via integration with the chatbot assistant (Section~\ref{sec:llm}), enhancing transparency, traceability, and reproducibility within AI deployment workflows.

\begin{figure}[h]
    \centering
    \includegraphics[width=\linewidth]{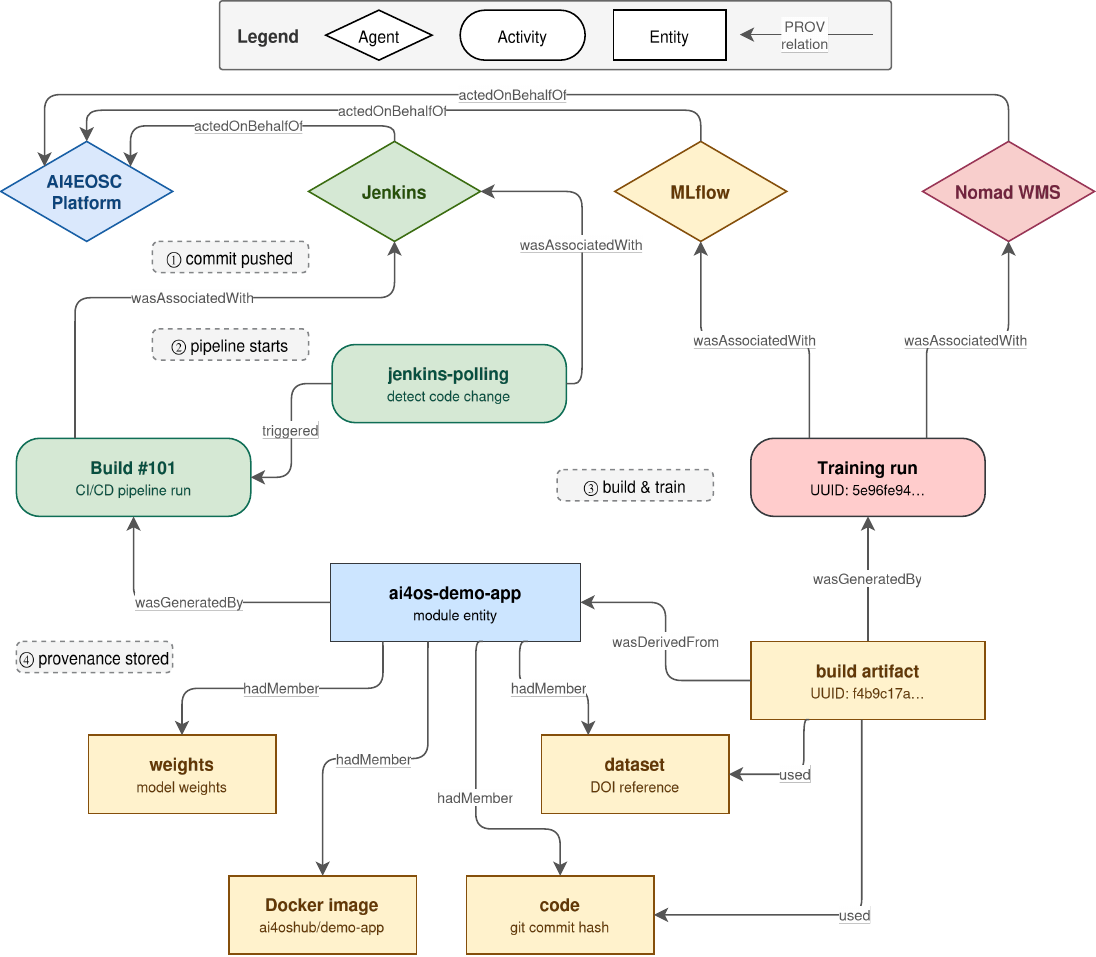}
    \caption{Example of a provenance graph built for a demo application.}
    \label{fig:provenance}
\end{figure}

Figure~\ref{fig:provenance} traces a concrete provenance record for the exemplary \texttt{ai4os-demo-app} module. When a developer commits new code, the Jenkins CI/CD agent detects the change (\texttt{jenkins-polling}), triggering Build 101. This activity, that is associated with Jenkins acting on behalf of the AI4EOSC platform, validates metadata, builds a Docker image, and notifies the provenance system. Simultaneously, a training run (UUID \texttt{5e96fe94}) executes on Nomad resources with metrics logged to MLflow, producing a versioned build artifact. All these relationships are captured as W3C PROV predicates (e.g. \texttt{wasGeneratedBy}, \texttt{wasAssociatedWith}, etc.), forming a complete, queryable lineage graph for the module.

\subsubsection{Openness}
\label{sec:openness}

AI4EOSC integrates open-source principles at its core: all the platform software is publicly available\footnote{\url{https://github.com/ai4os}}, which combined with the provided TOSCA deployment recipes used in the PaaS (\ref{sec:paas-orchestration}), enables communities to redeploy and built on top of our platform.

In a similar spirit, every AI module built with the platform has a public code repository attached to it, hosted under a common GitHub organization\footnote{\url{https://github.com/ai4os-hub}}. The code, as as well as any other downloadable asset (model weights, training dataset, metadata, provenance file, etc) is clearly linked from the AI module page in the AI4EOSC Dashboard (\ref{sec:papi}). To avoid vendor lock-in, AI4EOSC only allows but \textit{facilitates} the redeployment of all AI modules in external clouds (\ref{sec:external-deployment-platforms}).

\subsection{Platform extensibility and interoperability}
\label{sec:extensibility}

One of the design decisions of AI4EOSC is that any scientific platform must be able to adapt to the needs of  their user community, and not viceversa. In order to be able to comply with this design motto, our systems are built with extensibility in mind, meaning that we are able to integrate, to the technical possible extent, with external tools easily. In this section we present the three AI4EOSC extensibility and interoperability axes.

\subsubsection{External deployment platforms}
\label{sec:external-deployment-platforms}

All executable assets in the platform are based on Docker containers, that are being continuously released via reproducible builds into a public Docker Hub organization (\ref{sec:cicd}). This means that  the AI modules (\ref{sec:ai-modules}) are easily deployable at external cloud providers and in any other infrastructure where Docker is available. From the Dashboard (\ref{sec:dashboard}), we give users direct access to deploy in any external cloud provider via the Infrastructure Manager, that is part of our orchestration system (\ref{sec:paas-orchestration}). We have also developed connectors to be able to deploy on the EOSC EU Node, allowing users to leverage its resources.

\subsubsection{External AI catalogs}
\label{sec:external-ai-catalogs}

Due to our AI modules design (\ref{sec:ai-modules}), which hides the module internal workings behind a unified API \cite{garcia2019deepaas}, we are able to easily build connectors with external AI module catalogs. These connectors allow to deploy those external modules in our platform. To demonstrate this, we have built a connector to the BioImage Model Zoo catalog \cite{ouyang2022bioimage}, that leverages their standardized metadata to map their functionality into our platform, and making their AI modules directly available in our Dashboard (\ref{sec:dashboard}). This feature allows one-click deployments of those external modules, allowing the BioImage Model Zoo users to continue working with their tools, but making their outcomes available in the AI4EOSC marketplace.

\subsubsection{External datasets}
\label{sec:external-datasets}

AI/ML models rely heavily on data for development and training. Therefore, providing users with the ability to directly download data is an important feature. Additionally, in a metadata-rich platform like AI4EOSC, users can specify which dataset has been used to train a given model by referencing their DOI or URL. This capability is particularly valuable because it makes it easy to reproduce results, ensuring transparency and reliability in the research process. Our dashboard enables users to designate the data they wish to synchronize at launch time, streamlining the data management process.

We implement this functionality transparently via WMS sidecar tasks (\ref{sec:workload-management}) and extending the the DataHugger tool \cite{datahugger} with additional integrations, currently including Zenodo, Data Europa, Dryad or SeaNoe.

%   In addition to the DOI input, the Dashboard (\ref{sec:dashboard}) also offers a
%   custom built-in search functionality for Zenodo datasets.

\subsubsection{External storages}
\label{sec:external-storages}

In addition to the platform storage, we are able to connect user deployments with any RCLONE-compatible external storage, including Nextcloud, S3, Owncloud, Dropbox, Google Drive, etc. In order to do so, we allow users to easily link the platform with selected storage systems (e.g., NextCloud) or manually provide their user credentials for other storage providers.

\subsection{Security and access control}
\label{sec:security}

Taking into account the multi-user, multi-institution and multi-country nature of the AI4EOSC platform, security and data protection have been carefully considered in the design and implementation. In this regard, the platform adopts a layered security model that addresses authentication, authorization, secret management, tenant isolation, and data protection.

\subsubsection{Authentication and authorization}

The platform enforces authentication and authorization through dedicated Keycloak instance, using OpenID Connect (OIDC) as protocol. This way, all AI4EOSC services use this endpoint as the authoritative source of authentication, acting as a single trusted broker with external federated identity providers, such as the EGI Check-In or the MyAccessID (eduGAIN) services. This design, aligned with the AARC Blueprint Architecture \cite{aarc-2_project_aarc_2017}, reduces the trust boundary to a single managed service, reducing the complexity of dealing with identity providers at the individual service level, thus reducing the exposure level.

Authorization follows a role based access control (RBAC) derived both from the internal group mapping in the system and from the inherited entitlements from trusted identity providers. The platform services enforce different tiered access levels (from short-lived and constrained demo access to complete resource access).

\subsubsection{Secret management}

Besides user authentication, the platform manages a set of other secrets that can grant access to different components of the platform. These include storage tokens used to sync with external providers or tokens to join a federated learning scenario, among others. In order to manage them security, AI4EOSC enforces the usage of Vault, so that secrets are never exposed as plaintext to the platform infrastructure, as they are injected via the WMS (\ref{sec:workload-management}).

AI4EOSC does not enforce credential rotation, but users can define the secret lifetime when creating them and secrets can be revoked at any time via the platform dashboard or the API. This allows, for instance, to invalidate a token of a malicious client participating in a federated learning scenario or to revoke a compromised credential. This authentication is also intended to prevent poisoning attacks by unauthorized clients attempting to connect to the training, as will be further explored in the Section~\ref{sec:privacy_fl}.

\subsubsection{Tenant isolation}

AI4EOSC provides two levels of isolation.  At the organizational
level, Nomad namespaces are leveraged to partition resources across dedicated platform instances, for example for different research groups, institutions, or projects such as iMagine or KMD4EOSC (\ref{sec:adoption}). This ensures that the resources, catalogs, and configurations of different communities are fully segregated. Moreover, at workload level, user deployments are isolated at the internal network level, ensuring that deployments from different users do not interfere with, or observe, each other. Network-level access between deployments is restricted by default, so that inter-deployment communication is only possible through the explicitly exposed public endpoints, protected by the token-based access control described before.

\subsection{Privacy considerations}
\label{sec:privacy}

For a scientific platform like AI4EOSC, privacy is a paramount concern, as researchers frequently handle sensitive, proprietary, or geographically restricted datasets (e.g. medical datasets). Ensuring robust privacy mechanisms not only guarantees compliance with European data protection regulations such as the GDPR, but also establishes the necessary trust among institutions to enable collaborative AI workflows without compromising data confidentiality.

\subsubsection{Data retention}

According to the GDPR, the platform has been designed following the data minimization platform. Uploaded data is only stored in the user personal space and it is not retained after the user deletes it. When a user deployment is terminated, all locally cached data in the compute nodes (including datasets synchronized from external data providers) is completely removed. Users are required to adhere to the AI4EOSC privacy policy upon registration, and data processing agreements are in place with all federated infrastructure providers.

\subsubsection{Logging and traceability}

All user-facing actions are always routed through the Platform API (PAPI) (Section~\ref{sec:papi}), which provides a centralized and structured audit log of all platform activity. This includes job submissions, deployment lifecycle events, secret creation and revocation, and catalog interactions, enabling operators to trace the origin of any platform action.

\subsubsection{Federated learning}
\label{sec:privacy_fl}

Federated learning is a privacy-preserving approach by design, as it does not require participating clients to share their data with each other or with a central server. However, additional security and privacy risks remain, including poisoning attacks and client-side inference attacks, which must be addressed at the platform level.

As described in Section~\ref{sec:federated-learning-servers}, AI4EOSC extends Flower with token-based authentication and client weight divergence monitoring to defend against poisoning from both unauthorized and malicious authenticated clients. Server-side differential privacy and metric privacy \cite{SAINZPARDODIAZ2026115993} further protect against inference attacks on the aggregated model. Together, these mechanisms form a layered defense that operates transparently within the platform without requiring additional configuration from researchers. More specific measures remain complex to implement generically at the platform level and need to be addressed on a case-by-case basis.

\subsection{Reliability, Fault Tolerance, and Elasticity}
\label{sec:reliability}

The distributed and federated nature of the AI4EOSC platform requires a robust and reliable resource management system. Our design prioritizes system graceful degradations to maintain service availability despite the heterogeneous reliability of the underlying higly distributed provider nodes.

\subsubsection{Fault Tolerance and Self-healing}

To address potential failure of nodes in the WMS, we leverage the native self-healing capabilities of the underlying orchestrator (i.e., HashiCorp Nomad, as described in Section~\ref{sec:workload-management}).
The system maintains a constant state and heart-beat monitor of all federated nodes and supports graceful degradation due to the Nomad server redundancy. Additionally, Nomad incorporates consensus protocols to elect new cluster leaders in the event of a failure or network partitions.

\subsubsection{Resource Elasticity and Backpressure}

The platform differentiates between the elasticity requirements of training and inference workloads:

\begin{itemize}
    \item \textbf{AIaaS Autoscaling:} For inference services, the platform implements reactive horizontal autoscaling. Based on real-time metrics (e.g., request latency or CPU/GPU utilization), the AIaaS layer can dynamically provision additional container instances to handle traffic spikes, ensuring low-latency responses for end-user applications.

    \item \textbf{WMS Queueing and Determinism:} For training workloads, we adopt a deterministic resource allocation strategy. Given the high cost and limited availability of GPU resources across the scientific resource providers, we do not implement automated hardware provisioning (cluster autoscaling). Instead, we manage high demand through a queueing system. This prevents the over-saturation of the system as new jobs are held in a pending state until sufficient resources are released by completed tasks.
\end{itemize}

\section{Performance evaluation}
\label{sec:performance}

\subsection{Scalability}
\label{sec:performance-scalability}

\begin{figure}[ht]
    \centering
    \includegraphics[width=\linewidth]{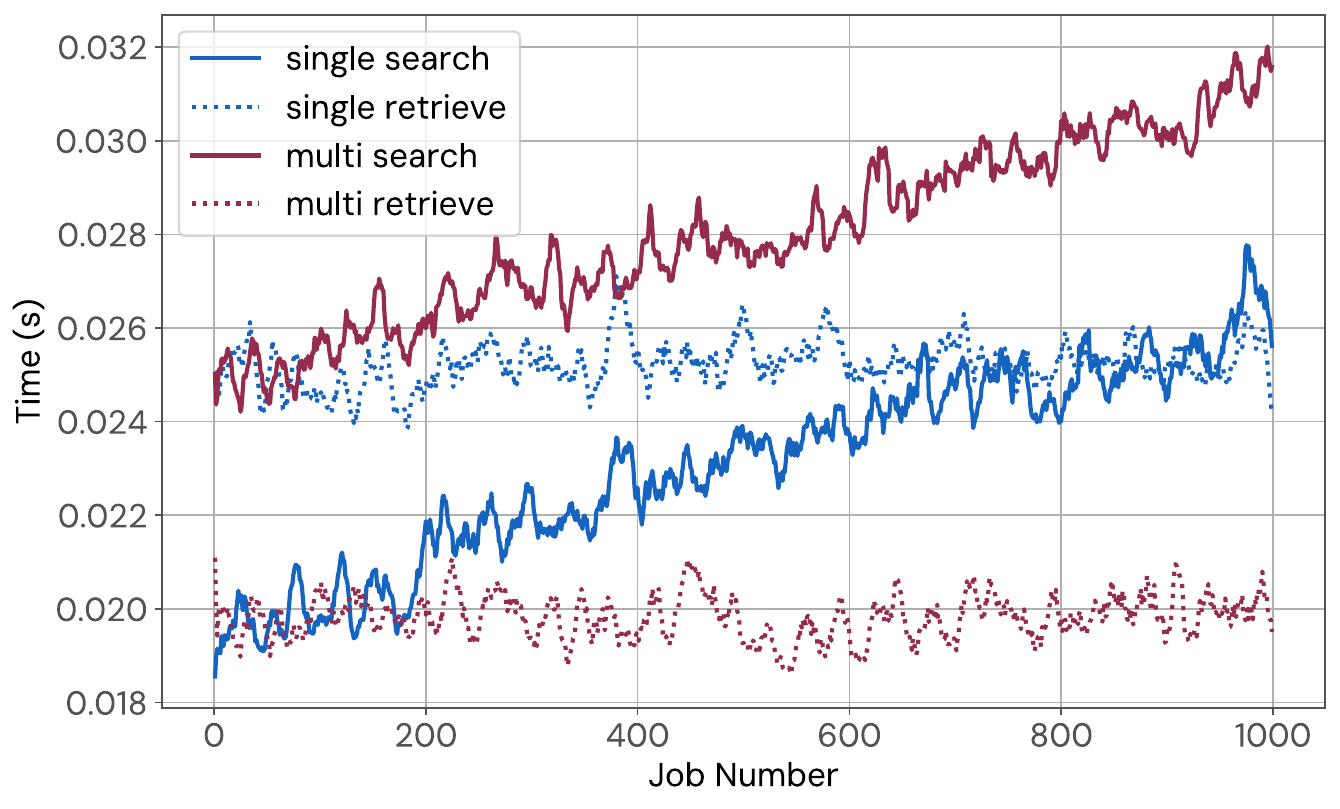}
    \caption{Search and retrieval times for a single WMS deployment, as function of the total jobs in the cluster. In blue, when launching all the jobs in the same datacenter. In red, when distributing the load equally among all datacenters. For clarity, a rolling mean average with window size 10 has been applied to smoothen the curves.}
    \label{fig:search-retrieve}
\end{figure}

To asses the scalability of the platform, in Figure~\ref{fig:search-retrieve} we show the time it takes to search and retrieve the full information of a single job, as a function of the number of the total jobs deployed in the AI4EOSC cluster federation. In blue we show the scenario when all the jobs are running in the same datacenter, while in red we show the case where the jobs are equally distributed across all the datacenters of the federation. As can be seen, search times increase sub-linearly with the number of total jobs, while job retrieval times stay constant, both in the range of few tens of milliseconds. More importantly, increasing the number of datacenters in the federation only increases search times by a small fixed cost. This, together with the sublinear increase in search times, shows the robust scalability capacity of the AI4EOSC federation, both in the number of jobs managed, as well as in the number of integrated infrastructure providers.

\subsection{Training Efficiency}
\label{sec:performance-efficiency}

To test the possible overhead of training in an AI4EOSC environment, we train the same Computer Vision model (from a real-life AI4EOSC usecase published in \cite{sainz2024personalized}) in three different scenarios: (1) directly in a Virtual Machine (VM); (2) inside a Docker container deployed in a Virtual Machine; and (3) in an AI4EOSC Development Environment (\ref{sec:development-environment}), which is, in the end, a Docker container running in a VM with Nomad/Consul orchestration (\ref{sec:workload-management}) and accessed via VScode (\ref{sec:development-environment}).

In each scenario, the model is trained for 25 epochs on the same resources (one NVIDIA T4 16GB GPU, 20 CPU cores, 86 GB RAM, 40 GB disk) with the same environments (Python 3.13, TensorFlow 2.20, CUDA 12.4, cuDNN version 9.21.01). The dataset is roughly 1 GB in size.

\begin{table}[ht]
    \centering
    \begin{tabular}{ccc}
    \toprule
        \textbf{VM} & \textbf{VM + Docker} & \textbf{AI4EOSC} \\
        \midrule
        $112.4 \pm 1.7$ & $112.6 \pm 1.2$  & $113.5 \pm 1.2$ \\
        \bottomrule
    \end{tabular}
    \caption{Training time (in seconds) for a real life AI4EOSC usecase trained for 25 epochs, averaged over 5 training runs.}
    \label{tab:efficiency-individual-training}
\end{table}

As shown in Table \ref{tab:efficiency-individual-training}, the dockerization on top of the VM does not add a significant difference in training times. Running in an AI4EOSC environment does result in a minor time increase (less than 1\%). This very small difference, caused by the Nomad/Consul orchestration overhead, demonstrates that the AI4EOSC platform is able to transparently serve federated resources at scale to users without significant performance loss.

\subsection{Federated Learning Efficiency}

Concerning the Flower FL server and the efficiency of the implementation in AI4EOSC, we have conducted four experiments with the same use case, considering in this set up the four clients presented in \cite{sainz2024personalized}. This experiments are as follows: (1) both the FL server and the four clients are deployed in AI4EOSC; (2) the FL server is deployed in AI4EOSC and the four clients run in four VMs; (3) both the FL server and the four clients run in five different VMs; (4) the FL server is deployed in a VM and the clients run in four separated AI4EOSC deployments. Concerning the training, each client trains the model for one epoch, and the FL process is conducted during five rounds. The results obtained in the four setups (mean and standard deviation after repeating the experiments five times) are shown in Table~\ref{tab:fl_time_results}.

\begin{table}[ht]
    \centering
    \begin{tabular}{lcc}
        \toprule
        \diagbox{\textbf{\textit{Server}}}{\textbf{\textit{Clients}}} &  \textbf{AI4EOSC} & \textbf{VMs}\\
        \midrule
        \textbf{AI4EOSC} & 229.18 $\pm$ 5.90 & 233.60 $\pm$ 14.61\\
        \textbf{VM} & 222.61 $\pm$ 14.16 & 226.96 $\pm$ 11.49\\
        \bottomrule
    \end{tabular}
    \caption{Training time (in seconds) for a federated learning setting with 5 rounds, averaged over 5 training runs with the standard deviation.}
    \label{tab:fl_time_results}
\end{table}

Regarding the client-side resources used in these scenarios, in AI4EOSC we deployed four instances, each configured with 4 vCPUS, 10 GB of RAM and 20 GB of disk memory. These specifications were chosen to ensure comparability with the selected flavor in the four VMs, matching the number of VCPS and disk capacity, and providing a similar RAM memory (10.7GB). In all the cases the same version of the FL server and the dependencies for running the clients were used (including Flower extensions).

As shown in Table~\ref{tab:fl_time_results}, running the FL server in AI4EOSC environment does not introduce a significant time overhead. Specifically, the difference compared to the best-case scenario (server on a VM and clients on AI4EOSC, with an average of 222.61 s) is less than 3$\%$ when the server runs on AI4EOSC with clients also on AI4EOSC, and does not exceed 5$\%$ in the worst-case scenario (server on AI4EOSC and clients on a VM). Furthermore, when the server runs on AI4EOSC, better results are obtained by running the clients in that environment as well, representing an improvement of nearly 2$\%$. However, this difference remains minimal in absolute terms. Furthermore, the best average performance is achieved when the server runs on a VM and the clients run on AI4EOSC (222.61 s). Nevertheless, considering the standard deviations, this result is very similar to the scenario in which both the server and the clients run on VMs within the same cloud infrastructure (226.96 s), with a difference of less than 2$\%$.

Finally, these results indicate that using the AI4EOSC platform in this FL setting does not introduce a significant computational overhead, with training times remaining comparable across configurations.

\subsection{Cold-start}
\label{sec:performance-cold}

Cold-start latency is observed mainly in the first invocations after service deployment or scale-up, when the platform must complete additional preparation steps such as image handling, container startup, and initialization before the request can be served.
Cold-start is particularly important for the AIaaS serverless component (based on OSCAR), as it handles short-lived inference calls (spawn container, make an inference, kill the container).
While execution times are application and container image-dependent, for a reference implementation, synchronous parallel OSCAR invocations complete with sub-second latency, indicating that the platform can sustain interactive request-response workloads when the service is already warm and resources are available. This was achieved in a 3-node OSCAR cluster (24 cores and 54 GBs of RAM available), showing an average first synchronous invocation time of 1521 ms and warm synchronous invocations of 992 ms.

Cold-start latency is not critical in the WMS, because the WMS is meant to executed longer jobs and tasks, like AI trainings, and thus user experience is not meaningfully impacted by warm/cold start delays (typically in the few seconds if the image is pre-cached in the node, to tens of seconds if the full image has to be pulled from the AI4EOSC container registry).

\section{Use cases and usage scenarios}
\label{sec:usage-scenarios}

\subsection{End-to-end user journey: precipitation nowcasting with federated learning}
\label{sec:user-journey}

To better illustrate how a researcher can interact with the AI4EOSC platform across the full AI/ML lifecycle, in this section we describe a concrete use case focused on radar-based precipitation nowcasting \cite{sainz2024personalized}. This use case aims to represent a broad class of scientific scenarios in which training data is distributed across institutions and cannot be centralized due to technical, legal, or administrative constraints.

\paragraph{Problem and data}
The goal of the selected case is to predict precipitation, expressed as Vertically Integrated Liquid (VIL), in a five minute horizon, using sequences of radar images captured by a meteorological radar located in the borderland between the Czech and Slovak Republic. The dataset covers four months of radar observations (between April and July 2016), with images captured every five minutes and stored as HDF5 files. To simulate a realistic federated scenario, the images are divided into four quadrants, each treated as a separate client with a distinct and heterogeneous data distribution. The goal of the case is to validate the feasibility of using federated learning for such predictions, where data is not directly accessible.

Data was stored in the storage system of the platform, that can be securely accessed from the user training jobs. In a realistic scenario, users would have been able to link their external storage system with their accounts, so that their datasets were not centrally stored, but rather accessed via their own storage systems only at training time. As outlined in Section \ref{sec:privacy}, data is not retained after the training process is finished.

\paragraph{Development}
The researchers deployed a Codespace via the AI4EOSC Dashboard, selecting a TensorFlow environment running on an NVIDIA Tesla T4 GPU (16 GB RAM, 8 CPU cores, 10 GB disk). The development environment provided secure and authenticated access to version-controlled repositories and the MLflow experiment tracking service, allowing the team to log and compare training runs across different model configurations and hyperparameter settings.

\paragraph{Federated training}
Once a suitable convolutional neural network architecture was identified and validated, the team configured a Flower federated learning server via the Dashboard, deploying four client instances within the platform itself, one corresponding to each zone. Token-based authentication was used to ensure that only authorized clients could participate in the training. The FL process ran for ten rounds of ten epochs each. Following the federated phase, a novel personalized FL approach (\textit{adapFL}) was applied, further fine-tuning the aggregated global model locally on each zone data for a set of ten additional epochs. This adaptive step allowed each client to recover zone-specific characteristics lost during federated aggregation.

\paragraph{Deployment and results}
In order to quickly validate the results, the generated models were deployed as permanent WMS endpoints for prototyping and evaluation, and as serverless AIaaS endpoints via OSCAR for on-demand inference. The platform CI/CD pipeline automatically built and published the corresponding Docker images, validated the module metadata, and triggered provenance graph updates, ensuring full traceability of the training process without any additional intervention from the researchers. The \textit{adapFL} approach yielded improvements in terms of both metrics evaluated (MSE and MAE) over individual training and vanilla FL in all four zones, demonstrating that privacy-preserving collaborative training on heterogeneous distributed data can yield better predictive models than local training alone.

\subsection{Platform adoption}
\label{sec:adoption}

Besides building a generic and horizontal platform, and in order to foster the adoption of the technology developed within multidisciplinary domains, we have closely worked with several initiatives. The aim is for these communities to collaborate on the co-design of the platform in order to build a customized version that adapts to the needs of the user community. The interactions with those communities has led to a virtuous circle, where the platform has guided them on their adoption of AI and they have in turn provided feedback for a user-driven platform development.

\subsubsection{iMagine}
The iMagine\footnote{\url{https://www.imagine-ai.eu/}} initiative is an European Union–funded research project that focuses on developing and operating an AI-enabled imaging platform for aquatic science. It aims to give researchers involved in the study of oceans, seas, coastal and inland waters access to rich image datasets, advanced AI-powered analysis tools and best-practice guidance for scientific image processing. The iMagine platform is built on top of the software stack developed within the AI4EOSC project.

Throughout the run of the iMagine project (which formally concluded in August 2025), the Competence Center team provided support to the users and collaborated closely with the developers of the AI4EOSC project, providing feedback to enhance the platform. This partnership enabled hands-on experience across the full AI workflow, from image collection and dataset preparation to model training and deployment for aquatic science applications, ultimately resulting in the publication \cite{AZMI2025103306}. It is important to note that the project has developed several use cases built on top of the services provided within the platform, including \cite{GARCIADIAZ2025103328} (focusing on satellite-derived bathymetry mapping) and \cite{11105453} (focusing on distance classification of marine vessels) among others.

\subsubsection{KMD4EOSC}

KMD4EOSC\footnote{\url{https://kmd4eosc.pl/}} is a Polish national project for data storage and sharing, and efficient processing of large volumes of data in HPC, BigData, and AI Models. The project is co-funded by European Funds for a Modern Economy Program and the European Union. The project is one of the flagship projects of the Polish Research Roadmap.

The KMD4EOSC project designs, builds and implements infrastructure, services and applications for storing and collecting scientific and economic data, as well as for processing and analyzing them and making them available to the scientific community and the economy in an efficient manner and in accordance with good practices, requirements and regulations. AI4EOSC has been deployed within the project to serve as a platform for providing the whole lifecycle of AI/ML models, integrated with repositories, datasets and lower infrastructure and data management services.

\subsubsection{AI4Life}
AI4Life\footnote{\url{https://ai4life.eurobioimaging.eu/}}, a Horizon Europe funded project to empower life science researchers to harness the full potential of AI/ML methods for bio-image analysis, began collaborating with AI4EOSC in order to leverage the AI4EOSC software stack and platform. The AI4EOSC and AI4Life teams worked on both platforms integration, as a means to implement and pilot a model for interoperability across research infrastructures, aligned with FAIR principles. This integration enables efficient sharing and use of AI/ML models and computing resources across domains.

Central to this effort is deploying models from the BioImage Model Zoo within the AI4Life cloud marketplace. Compatibility was ensured through defined execution environments such as PyTorch, standardized metadata rich input and output formats, and consistent prediction pipelines, allowing deployment via Dockerfile and DEEPaaS compatible APIs.

\section{Conclusion and Future Work}
\label{sec:conclusion}

The growing adoption of AI and ML in scientific research demands platforms that go beyond industrial MLOps frameworks to address the specific needs of open science:  reproducibility, FAIR compliance, provenance tracking, and seamless integration with federated research infrastructures to enable easy access to the technologies by researchers. This paper has presented AI4EOSC, a federated, open-source platform that operationalizes these principles across the full AI/ML lifecycle within the European Open Science Cloud ecosystem.

The main architectural contribution of AI4EOSC is not the provision of any single or standalone tool, but rather the integration of development, training, deployment, and provenance into a unified, FAIR-by-design system that enforces open science compliance at the infrastructure level, alleviating the burden from individual researchers. AI4EOSC main innovations include a lightweight inter-cluster federation based on Consul and Nomad that avoids the infrastructure fragmentation and silos of Kubernetes-centric stacks, an
automated W3C PROV-compliant provenance pipeline that captures full model lineage with limited user intervention, and extended federated learning support with token-based authentication, novel aggregation strategies, and server-side differential privacy. AI4EOSC value proposition has been validated through three real-world community adoptions: iMagine, KMD4EOSC, and AI4Life. This demonstrates consistent functionality and deployment across heterogeneous resource providers and diverse scientific domains.

Nevertheless several limitations need to be acknowledged. Firstly, the current provenance model, while comprehensive for single-model workflows, requires further work to handle multi-model inference pipelines and federated learning scenarios where data remains siloed
across institutions. Secondly, although the platform currently supports semantic interoperability through MLDCAT-AP serialization, full machine-actionability of AI/ML assets across heterogeneous repositories and data spaces remains an open and evolving challenge: deeper integration with emerging standards and external catalogs is needed to allow assets to be not only findable and accessible, but autonomously actionable by
external systems. Lastly, the operational overhead of maintaining a geographically distributed federation (including failure handling, performance variability across heterogeneous providers, and long-term governance) represents a practical challenge that will grow as the federation expands. Although automation is in place in order to alleviate
this fact, this is a common issue of any geographically and complex distributed infrastructure, and AI4EOSC alleviates this fact by leveraging a single and unified control plane.

Future work in the near term will be focused on deepening semantic interoperability and machine actionability by extending the current support for MLDCAT-AP, and focusing on
emerging standards such as Croissant-ML \cite{akhtar_croissant_2024}. Moreover, we will implement the OAI-PMH protocol to enable automated metadata harvesting across external marketplaces and repositories. We will also extend the MLflow experiment tracking component to serve as a full model registry, enriching stored models with semantic provenance information, as we are already doing with the modules published through the dashboard. In the medium term, we plan to address the provenance gaps identified above, in particular for federated and multi-model scenarios, and to strengthen the platform's energy and sustainability awareness by integrating fine-grained carbon and water footprint measurements via Wattnet \cite{castrillo_wattnet}, implementing platform level optimizations, and informing users about the impact of their workloads. In the longer term, we aim to grow the current federation by lowering the barrier for new resource providers to join, and to conduct systematic user studies to measure the actual impact of the platform on researcher productivity and reproducibility outcomes.

\section{Acknowledgments}

The authors acknowledge Marcos Lloret Iglesias, the AI4EOSC project manager, for his constant support and help towards bringing the project to fruition.
The authors acknowledge the funding and support from the AI4EOSC project ``Artificial Intelligence for the European Open Science Cloud'', that has received funding from the European Union's Horizon Europe research and innovation programme under grant agreement number 101058593. ALG also acknowledges the support from the \textit{Consejería de Educación, Formación profesional y Universidades} of the \textit{Gobierno de Cantabria} via the ``\textit{Actividad estructural para el desarrollo de la investigación del Instituto de Física de Cantabria}'' project.

\bibliographystyle{elsarticle-num-names}
\bibliography{references.bib}

\end{document}